\documentclass[a4paper,fleqn,usenatbib]{mnras}

\usepackage[T1]{fontenc}
\usepackage{ae,aecompl}

\usepackage{graphicx}	
\usepackage{amsmath}	
\usepackage{amssymb}	

\title[Abundance gradients]{Abundance gradients along the Galactic disc from chemical evolution models}

\author[Grisoni et al.]{V. Grisoni$^1$\thanks{E-mail: grisoni@oats.inaf.it}, E. Spitoni$^{1,4}$, F. Matteucci$^{1, 2, 3}$\\
 $^1$ Dipartimento di Fisica, Sezione di Astronomia, Universit\`a di Trieste, via G.B. Tiepolo 11, I-34131, Trieste, Italy \\  
 $^2$ I.N.A.F. Osservatorio
  Astronomico di Trieste, via G.B. Tiepolo 11, I-34131, Trieste,
  Italy\\
 $^3$ I.N.F.N. Sezione di Trieste, via Valerio 2, 34134 Trieste, Italy\\
 $^4$ Stellar Astrophysics Centre, Department of Physics and Astronomy, Aarhus University, Ny Munkegade 120, DK-8000 Aarhus C,
\\Denmark
}

\begin{document}
\date{Accepted . ; in original form xxxx}

\pagerange{\pageref{firstpage}--\pageref{lastpage}} \pubyear{xxxx}

\maketitle

\label{firstpage}

\begin{abstract}
In this paper, we study the formation and chemical evolution of the Milky Way disc with particular focus on the abundance patterns ([$\alpha$/Fe] vs. [Fe/H]) at different Galactocentric distances, the present-time abundance gradients along the disc and the time evolution of abundance gradients. We consider the chemical evolution models for the Galactic disc developed by Grisoni et al. (2017) for the solar neighborhood, both the two-infall and the one-infall ones, and we extend our analysis to the other Galactocentric distances. In particular, we examine the processes which mainly influence the formation of the abundance gradients: the inside-out scenario, a variable star formation efficiency, and radial gas flows. We compare our model results with recent abundance patterns obtained along the Galactic disc from the APOGEE survey and with abundance gradients observed from Cepheids, open clusters, HII regions and PNe. We conclude that the inside-out scenario is a key ingredient, but cannot be the only one to explain abundance patterns at different Galactocentric distances and abundance gradients. Further ingredients, such as radial gas flows and variable star formation efficiency, are needed to reproduce the observed features in the thin disc. The evolution of abundance gradients with time is also shown, although firm conclusions cannot still be drawn.
\end{abstract}


\section{Introduction}

In order to study the formation and chemical evolution of our Galaxy, a fundamental constraint is represented by abundance gradients along the Galactic thin disc. Furthermore, recent observational data of abundance patterns at various Galactocentric distances represent another important constraint for understanding the formation and evolution of the Milky Way disc. In this context, it is important to have several elements, produced by stars with different masses and timescales (Adibekyan et al. 2012; Bensby et al. 2014; Hayden et al. 2015; Magrini et al. 2017), to gain information about the nucleosynthesis channels, sites of production and timescales of enrichment of each chemical element, but also about the star formation history of the Galactic disc.
\\Abundance gradients have been observed in many spiral galaxies and show that the abundances of metals decrease outward from the Galactic center. Generally, a good agreement between observational properties of the Galaxy and model predictions is obtained by assuming that the disc formed by infall of gas (Chiosi 1980; Matteucci and Francois 1989; Ferrini et al. 1994; Chiappini et al. 1997, 2001; Cescutti et al. 2007; Colavitti et al. 2009; Chiappini 2009; Magrini et al. 2009; Spitoni and Matteucci 2011; Mott et al. 2013; Haywood et al. 2013; Snaith et al. 2015; Kubryk et al. 2015a,b; Prantzos et al. 2018).  In particular, a good assumption for reproducing abundance gradients is that the timescale for the formation of the Galactic thin disc increases with Galactocentric radius according to the inside-out scenario (Matteucci and Francois 1989; Chiappini et al. 2001). Cescutti et al. (2007) showed that a two-infall model with inside-out scenario gives a very good agreement with the data of Cepheids for many elements (Andrievsky et al. 2002a,b,c, 2004; Luck et al. 2003). Colavitti et al. (2009) showed that it is fundamental to assume an inside-out scenario, but also other ingredients, such as a threshold in the gas density for the star formation rate or a variable star formation efficiency (higher in the inner region than in the outer ones), in order to reproduce the present day gradients in the outer disc. More recently, Pilkington et al. (2012) have supported the conclusion that spiral discs form inside-out. To mantain consistency with the dynamical consequence of infall, also radial gas flows have to be taken into account (Spitoni and Matteucci 2011; Bilitewski and Sch{\"o}nrich 2012; Wang and Zhao 2013; Spitoni et al. 2013; Mott et al. 2013; Cavichia et al. 2014; Pezzulli et al. 2017). The infalling gas has a lower angular momentum than the circular motions in the disc, and mixing with the gas in the disc induces a net radial inflow. Lacey and Fall (1985) found that the gas inflow velocity (v$_R$) is up to a few km s$^{-1}$ and at 10 kpc is v$_R$=-1 km s$^{-1}$. Goetz and Koeppen (1992) developed numerical and analytical models including radial gas flows and they concluded that radial flows alone cannot explain the abundance gradients, but they are an efficient process to amplify the existing ones. Portinari and Chiosi (2000) implemented radial gas flows in a detailed chemical evolution model characterized by a single infall episode. More recently, Spitoni and Matteucci (2011) and Spitoni et al. (2013) have taken into account inflows of gas in detailed one-infall models for the Milky Way and M31 respectively, treating the evolution of the thin disc independently from the halo and thick disc. Spitoni and Matteucci (2011) tested also the radial flows in a two-infall model, but only for oxygen. They found that the observed gradient of oxygen can be reproduced if the gas inflow velocity increases in modulus with the Galactocentric distance, in both the one-infall and two-infall models. A similar approach was followed also by Mott et al. (2013), who studied also the evolution with time of the gradients. At variance with the previous papers where the velocity patterns of the inflow were chosen to produce a best-fit model, Bilitewski and Sch{\"o}nrich (2012) presented a chemical evolution model where the flow of gas is directly linked to physical properties of the Galaxy like the angular momentum budget. The resulting velocity patterns of the flows of gas are time dependent and show a non linear trend, always decreasing with decreasing Galactocentric distance. At a fixed Galactocentric distance, the velocity flows decrease with time.
\\The time evolution of abundance gradients has been studied in several works and in literature various predictions have been made by chemical evolution models. Some authors predicted that the gradient steepens with time (Chiappini et al. 2001; Mott et al. 2013), whereas others suggested that the gradient flattens in time (Prantzos and Boissier 2000; Moll{\'a} and D{\'{\i}}az 2005; Vincenzo et al. 2018; Minchev et al. 2018). The discrepancy between different model predictions is due to the fact that chemical evolution is very sensitive to the prescriptions of the physical processes that lead to the differential enrichment of inner and outer discs, and the flattening or steepening of gradients with time depends on the interplay between infall rate, star formation rate along the disc and also on the presence of a threshold in the gas density for the star formation (Kennicutt 1998a,b). Different recipes of star formation or gas accretion mechanisms can provide different abundance gradients predictions. From the observational point of view, there have been some studies to infer the time evolution of gradients from planetary nebulae (PNe) of different ages (Maciel and Costa 2009, 2013; Stanghellini and Haywood 2010, 2018). In particular, Maciel and Costa (2009) found a time flattening of the gradients during the last 6-8 Gyr, whereas Maciel and Costa (2013) concluded that the radial gradient has not changed appreciably during the Galactic lifetime. On the other hand, Stanghellini and Haywood (2010) found that the Galactic PN gradient may steepen with Galaxy evolution, and this conclusion has been confirmed by Stanghellini and Haywood (2018). Xiang et al. (2015) studied the evolution of stellar metallicity gradients of the Milky Way disk from main sequence turn-off stars from LAMOST Spectroscopic Survey of the Galactic Anticentre (LSS-GAC), and concluded that the radial gradients, after being essentialy flat at the earliest epochs of disc formation, steepen with time, reaching a maximum at age 7-8 Gyr, and then they flatten again, suggesting a two-phase disc formation history (see also Huang et al. 2015, Xiang et al. 2017). Furthermore, Anders et al. (2017) measured the age dependence of the radial metallicity distribution in the Galactic thin disc over cosmic time from CoRoT and APOGEE red giants, and concluded that the slope of the radial iron gradient was compatible with a flat distribution for older ages, then it steepens and finally flattens again. These results are in agreement with the one of the Geneva-Copenhagen survey (Nordstr{\"o}m et al. 2004; Casagrande et al. 2011), but there are differences with the LAMOST study of Xiang et al. (2015), possibly due to systematic shifts in the distance and age scales. Furthcoming data from asteroseismic and spectroscopic observations will be fundamental to further constrain the time evolution of the radial abundance gradients.
\\The aim of this paper is to study the abundance ratios and abundance gradients in the Galactic thin disc at the present time and their evolution on the basis of detailed chemical evolution models. We consider the recent chemical evolution models for the Galactic thin disc developed by Grisoni et al. (2017) for the solar neighborhood and we extend our analysis to other Galactocentric distances. In particular, we examine the processes which mainly influence the formation of abundance gradients: i) the inside-out scenario for the formation of the Galactic thin disc, ii) a variable star formation efficiency, and iii) radial gas flows along the Galactic disc.
\\This paper is organized as follows. In Section 2, we show the observational data which have been considered to make a comparison with the predictions of our chemical evolution models. In Section 3, we describe the chemical evolution models used in this work. In Section 4, we present the comparison between observations and model predictions. Finally, in Section 5, we summarize our results and conclusions.

\section{Observational data}

Abundance gradients can be studied by using several tracers, such as HII regions, planetary nebulae (PNe), Cepheids and open clusters (OCs).
\\In this paper, we adopt the Cepheids data by Luck and Lambert (2011) and Genovali et al. (2015), and the OCs data from Magrini et al. (2017). We adopt the HII region data by Deharveng et al. (2000), Esteban et al. (2005), Rudolph et al. (2006) and Balser et al. (2015), and the PNe data by Stanghellini and Haywood (2018).
Since PNe and OCs, due to their age spread, are not representative of a single epoch in the Galaxy lifetime, when comparing with the present-time gradient we consider only young PNe and young OCs; in particular, we consider only YYPNe, i.e. PNe whose progenitor stars are younger than 1 Gyr (Stanghellini and Haywood 2018), and also the young OCs with ages less than 1 Gyr of Magrini et al. (2017). The observational data are plotted with their typical errors (see references for further details on the typical errors both in abundances, distances, and ages as well). Regarding the systematic errors affecting the distance estimation of the used stellar populations, we note that the distance scale of PNe is much more uncertain of those of the other tracers.
\\For comparison with the time evolution of the radial metallicity gradient, we adopt the recent APOGEE data by Anders et al. (2017) and the PNe data by Stanghellini and Haywood (2018).
\\We also look at how the abundance patterns of [$\alpha$/Fe] vs. [Fe/H] vary with Galactocentric distance and we consider the APOGEE data of Hayden et al. (2015) for comparison with our models; in particular, among the $\alpha$-elements, here we focus on magnesium and we compare our predictions with the observed [Mg/Fe] vs. [Fe/H] at various Galactocentric distances.

\begin{table*}
\caption{Input parameters for the chemical evolution models. In the first column, we write the name of the model. In the second column, there is the star formation efficiency of the thin disc at different radii (4-6-8-10-12-14-16 kpc from the Galactic center). Finally, in the last column, we indicate the presence or the absence of radial flows. In each model, we adopt Kroupa et al. (1993) IMF.}
\label{tab_01}
\begin{center}
\begin{tabular}{c|cccccccccc}
  \hline
\\
 Model &$\nu$(4-6-8-10-12-14-16 kpc) & Radial flows\\
&[Gyr$^{-1}$]& & \\
\\
\hline

2IM A & const &  no \\


 \hline

1IM A & const &  no \\

 \hline

1IM B & 8.0-4.0-1.0-0.5-0.2-0.1-0.05 & no \\

 \hline

1IM C & const & yes \\

 \hline

1IM D & 8.0-4.0-1.0-0.5-0.2-0.1-0.05 & yes \\

 \hline

\end{tabular}
\end{center}
\end{table*}

\section{The models}

The chemical evolution models adopted here are the ones developed in Grisoni et al. (2017) for the solar neighborhood, that now we extend to the other Galactocentric distances. The models are as follows.
\begin{itemize}
\item The two-infall model (Chiappini et al. 1997, Romano et al. 2010) revisited and applied to the thick and thin discs. This model assumes that the discs form as a result of two main infall episodes: during the first infall episode, the thick disc formed, whereas during the second one a much slower infall of gas, delayed with respect to the first one, gives rise to the thin disc.
\item The parallel model, adopting two separate one-infall approaches for the thick and thin discs, respectively; in this model, we consider the thick and the thin disc stars as formed in two distinct evolutionary phases, which evolve independently.
\end{itemize}
In this work, the Galactic thin disc is approximated by several independent rings, 2 kpc wide, whereas the evolution of the thick disc evolves as a one-zone with radius of 8 kpc (see Haywood et al. 2018). The basic equations that follow the time evolution of $G_i$, i.e. the mass fraction of the element $i$ in the gas, are (see Matteucci 2012 for an exhaustive description):
\begin{align} \label{eq_01}
\begin{split}
& \dot G_i(r,t)= -\text{SFR}(r,t) X_i(r,t)+R_i(r,t)+ \dot G_i(r,t)_{inf}
\end{split}
\end{align}
where $\text{SFR}(r,t)$ is the star formation rate (SFR), $X_i(r,t)$ the abundance by mass of the element $i$, $R_i(r,t)$ refers to the rate of restitution of matter from the stars of different masses into the ISM, and $\dot G_i(r,t)_{inf}$ is the gas accretion rate.
\\The first term in the right hand side of Eq.\,\eqref{eq_01} ($\text{SFR}(r,t) X_i(r,t)$) corresponds to the rates at which the chemical elements are subtracted from the interstellar medium (ISM) to be included into stars. The SFR is parametrized according to the Schmidt-Kennicutt law (Kennicutt 1998a):
\begin{equation} \label{eq_03}
\text{SFR}(r,t)=\nu \sigma_{gas}^k(r,t),
\end{equation}
where $\sigma_{gas}$ is the surface gas density, $k=1.4$ is the law index and $\nu$ is the star formation efficiency (SFE), which is tuned to reproduce the present time SFR. The adopted initial mass function (IMF) is the Kroupa et al. (1993) one.
\\In the term $R_i(r,t)$ in the right hand side of Eq.\,\eqref{eq_01}, we take into account detailed nucleosynthesis from low and intermediate mass stars, Type Ia SNe (originating from white dwarfs in binary systems) and Type Ib, Ic and II SNe (originating from core-collapse massive stars). The contribution of Type Ia SNe was first introduced by Matteucci and Greggio (1986). Here, the rate is calculated in the framework of the single-degenerate scenario for the progenitor of these SNe, i.e. a C-O white dwarf plus a red giant companion; in Matteucci et al. (2009), it has been shown that this scenario is equivalent to the double degenerate one concerning the effects on Galactic chemical evolution. Here, we adopt the same nucleosynthesis prescriptions of Grisoni et al. (2017)
\\The last term in Eq.\,\eqref{eq_01} is the gas accretion rate. In particular, in the two-infall model the gas infall law is given by:
\begin{align} \label{eq_02}
\dot G_i(r,t)_{inf}=A(r)(X_i)_{inf}e^{-\frac{t}{\tau_1}}+B(r)(X_i)_{inf}e^{-\frac{t-t_{max}}{\tau_2}},
\end{align}
where $G_i(r,t)_{inf}$ is the infalling material in the form of element $i$ and $(X_i)_{inf}$ is the composition of the infalling gas which is assumed to be primordial. The parameter t$_{max}$ refers to the time for the maximum mass accretion onto the disc and roughly corresponds to the end of the thick disc phase. The parameters $\tau_1$ and $\tau_2$ represent the timescales for mass accretion in the thick and thin discs, respectively. These timescales are free parameters of the model and they are constrained mainly by comparison with the observed metallicity distribution function of long-lived stars in the solar vicinity. In the solar vicinity, Grisoni et al. (2017) found that the best values for these timescales are $\tau_1=0.1$ Gyr and $\tau_2=7$ Gyr. In this paper, we assume that the timescale for mass accretion in the Galactic thin disc changes with the Galactocentric distance according to the inside out scenario (see Chiappini et al. 2001):
\begin{align} \label{eq_02a}
\tau_2 \text{[Gyr]}=1.033 \text{r} \text{[kpc]}-1.267,
\end{align}
whereas the timescale of the thick disc is fixed and so there is no inside-out scenario for the thick disc (see also Haywood et al. 2018). The quantities $A(r)$ and $B(r)$ are two parameters fixed by reproducing the present time total surface mass density in the solar neighborhood. In particular, the surface mass density is equal to 65 M$_{\odot}$pc$^{-2}$ for the thin disc, and 6.5 M$_{\odot}$pc$^{-2}$ for the thick disc (Nesti and Salucci 2013). In this paper, we assume that the surface mass density in the Galactic thin disc changes with the Galactocentric distance according to:
\begin{align} \label{eq_02a}
\sigma(r)=\sigma_0 e^{-\frac{r}{r_D}},
\end{align}
where $\sigma_0$=1413 M$_{\bigodot}$ pc$^{-2}$ is the central total surface mass density and $r_D$=2.6 kpc is the scale lenght (Nesti and Salucci 2013). In the case of the thick disc, it is constant and equal to 6.5 M$_{\odot}$pc$^{-2}$ up to 8 kpc and then it decrease with the inverse of the Galactocentric distance (see Chiappini et al. 2001).
\\On the other hand, in the parallel model, since we assume two distinct infall episodes, the gas infall is described as:
\begin{equation} \label{eq_07}
(\dot G_i(r,t)_{inf})|_{thick}=A(r)(X_i)_{inf}e^{-\frac{t}{\tau_1}},
\end{equation}
\begin{equation} \label{eq_08}
(\dot G_i(r,t)_{inf})|_{thin}=B(r)(X_i)_{inf}e^{-\frac{t}{\tau_2}},
\end{equation}
for the thick disc and for the thin disc, respectively. The quantities $A(r)$ and $B(r)$ and the parameters $\tau_1$ and $\tau_2$ have the same meaning as discussed for Eq. \,\eqref{eq_02}.

\subsection{Implementation of radial inflows}

We implement radial inflows of gas in our chemical evolution models following the prescriptions of Spitoni and Matteucci (2011).
\\We define the $k$-th shell in terms of the Galactocentric radius $r_k$, where its inner and outer edge are labeled as $r_{k-\frac{1}{2}}$ and $r_{k+\frac{1}{2}}$. Through these edges, gas inflow occurs with velocity $v_{k-\frac{1}{2}}$ and $v_{k+\frac{1}{2}}$, respectively. The flow velocities are assumed to be positive outward and negative inward.
\\Radial inflows with a flux $F(r)$ alter the gas surface density $\sigma_{gk}$ in the $k$-th shell according to:
\begin{align} \label{eq_07}
\left[ \frac{d\sigma_{gk}}{dt} \right]_{rf}=-\frac{1}{\pi (r^2_{k+\frac{1}{2}}-r^2_{k-\frac{1}{2}})}[F(r_{k+\frac{1}{2}})-F(r_{k-\frac{1}{2}})],
\end{align}
where
\begin{align} \label{eq_08}
F(r_{k+\frac{1}{2}})=2\pi r_{k+\frac{1}{2}} v_{k+\frac{1}{2}} [\sigma_{g(k+1)}],
\end{align}
and
\begin{align} \label{eq_09}
F(r_{k-\frac{1}{2}})=2\pi r_{k-\frac{1}{2}} v_{k-\frac{1}{2}} [\sigma_{g(k-1)}].
\end{align}
We take the inner edge of the $k$-shell, $r_{k-\frac{1}{2}}$, at the midpoint between the characteristic radii of the shells $k$ and $k-1$, and similarly for the outer edge $r_{k+\frac{1}{2}}$:
\begin{align} \label{eq_10}
r_{k-\frac{1}{2}}=\frac{r_{k-1}+r_k}{2},
\end{align}
and
\begin{align} \label{eq_11}
r_{k+\frac{1}{2}}=\frac{r_{k}+r_{k+1}}{2},
\end{align}
We get that:
\begin{align} \label{eq_12}
(r^2_{k+\frac{1}{2}}-r^2_{k-\frac{1}{2}})=\frac{r_{k+1}-r_{k-1}}{2}(r_k+\frac{r_{k-1}+r_{k+1}}{2}).
\end{align}
Therefore, by inserting these quantities into Eq. \ref{eq_07}, we obtain the radial flow term to be added in Eq. \ref{eq_01}:
\begin{align} \label{eq_13}
\left[ \frac{d G_i(r_k,t)}{dt} \right]_{rf}=-\beta_k G_i(r_k,t)+\gamma_k G_i(r_{k+1},t),
\end{align}
where
\begin{align} \label{eq_14}
\beta_k=-\frac{2}{r_k+\frac{r_{k-1}+r_{k+1}}{2}}[v_{k-\frac{1}{2}} \frac{r_{k-1}+r_k}{r_{k+1}-r_{k-1}}],
\end{align}
and
\begin{align} \label{eq_15}
\gamma_k=-\frac{2}{r_k+\frac{r_{k-1}+r_{k+1}}{2}}[v_{k+\frac{1}{2}} \frac{r_{k}+r_{k+1}}{r_{k+1}-r_{k-1}}]\frac{\sigma_{(k+1)}}{\sigma_k},
\end{align}
where $\sigma_{(k+1)}$ and $\sigma_k$ are the present time total surface mass density profile at the radius $r_{k+1}$ and $r_k$, respectively. We assume that there are no flows from the outer parts of the disc where there is no star formation. In our implementation of the radial inflow of gas, only the gas that resides inside the Galactic disc within the radius of 16 kpc can move inward by radial inflow.
\\We adopt a variable velocity for the radial gas flows. In particular, the modulus of the radial inflow velocity as a function of the Galactocentric distance is assumed to be:
\begin{align} \label{eq_16}
|v_R|=\frac{R_g}{4}-1,
\end{align}
where the range of the velocities span the range 0-3 km s$^{-1}$, in accordance with previous works (Wong et al. 2004; Sch{\"o}nrich and Binney 2009; Spitoni and Matteucci 2011; Mott et al. 2013). Furthermore, our radial inflow patterns are in agreement with the ones computed by Bilitewski and Sch{\"o}nrich (2012), imposing the conservation of the angular momentum.

\begin{figure*}
\includegraphics[scale=0.58]{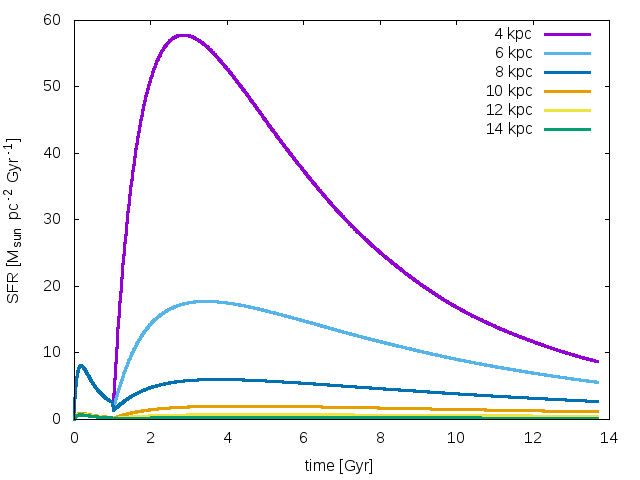}
 \caption{Time evolution of the SFR, as predicted by the model 2IM A at various Galactocentric distances.}
 \label{fig_01}
\end{figure*}

\begin{figure*}
\includegraphics[scale=0.35]{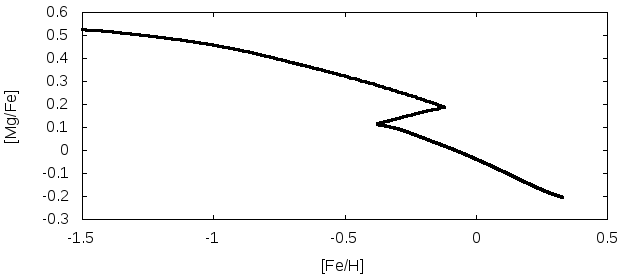}
\includegraphics[scale=0.35]{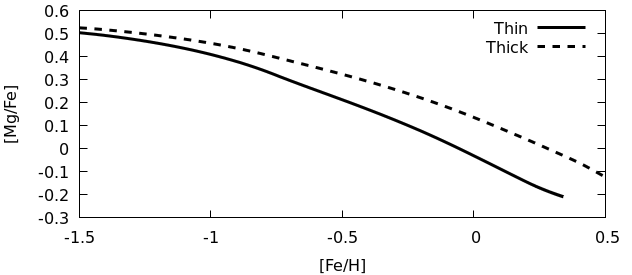}
 \caption{Predicted abundance patterns of [Mg/Fe] vs. [Fe/H] in the solar neighborhood for the two scenarios, the two-infall (on the left) and the parallel (on the right).}
 \label{fig_01b}
\end{figure*}

\begin{figure*}
\includegraphics[scale=0.35]{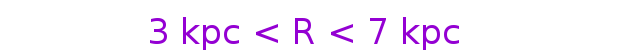}
\includegraphics[scale=0.35]{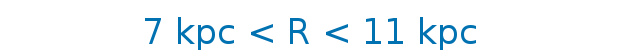}
\includegraphics[scale=0.35]{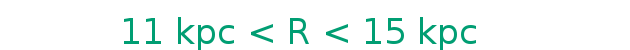}
\includegraphics[scale=0.35]{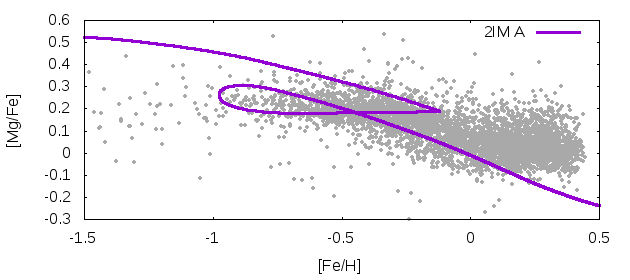}
\includegraphics[scale=0.35]{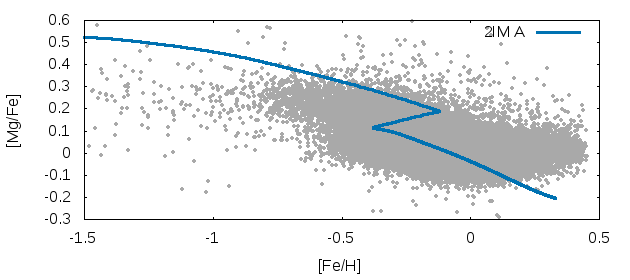}
\includegraphics[scale=0.35]{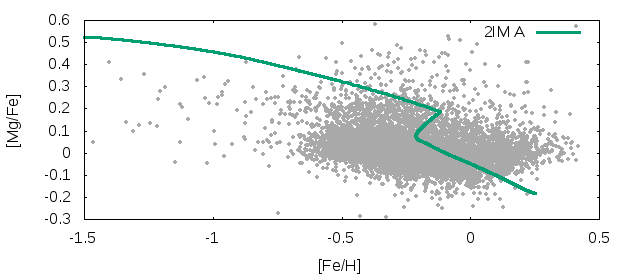}
\includegraphics[scale=0.35]{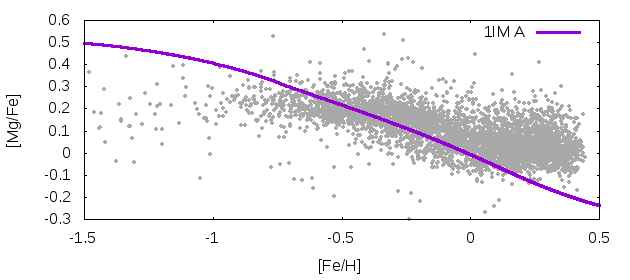}
\includegraphics[scale=0.35]{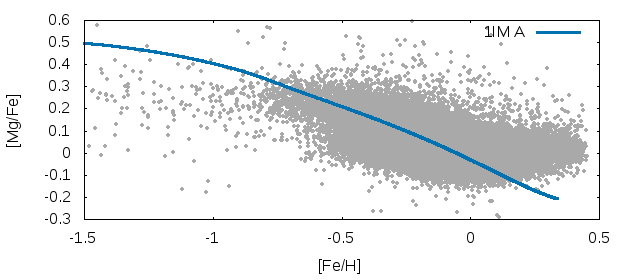}
\includegraphics[scale=0.35]{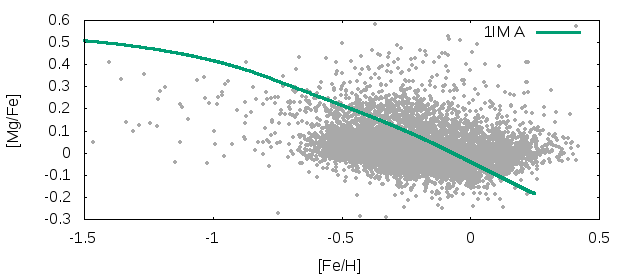}
\includegraphics[scale=0.35]{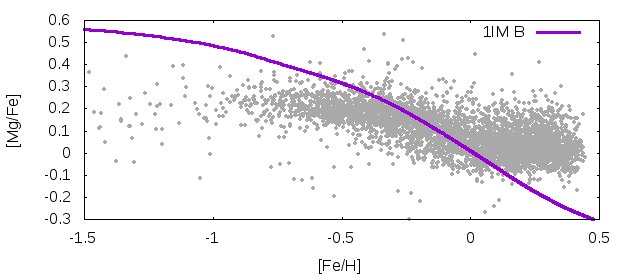}
\includegraphics[scale=0.35]{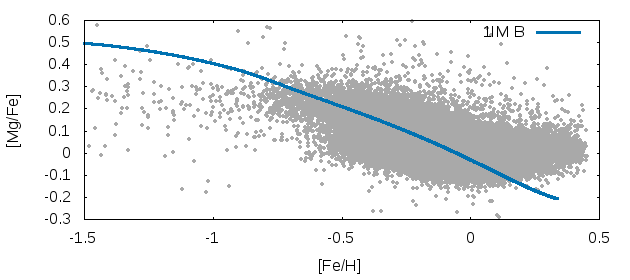}
\includegraphics[scale=0.35]{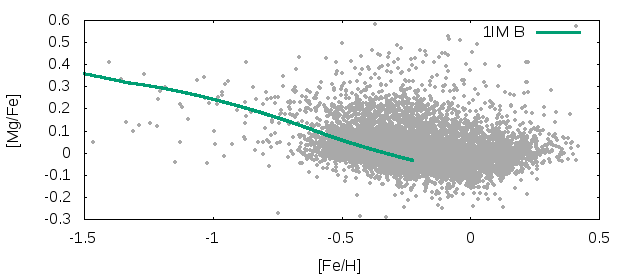}
\includegraphics[scale=0.35]{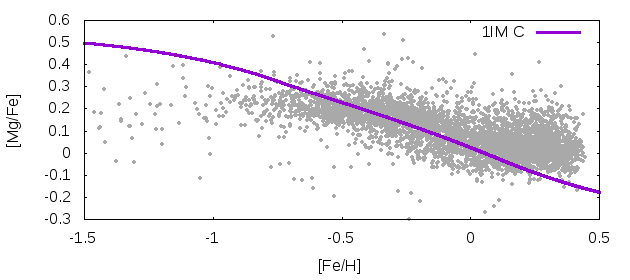}
\includegraphics[scale=0.35]{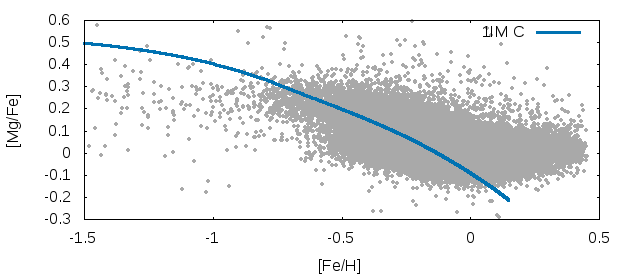}
\includegraphics[scale=0.35]{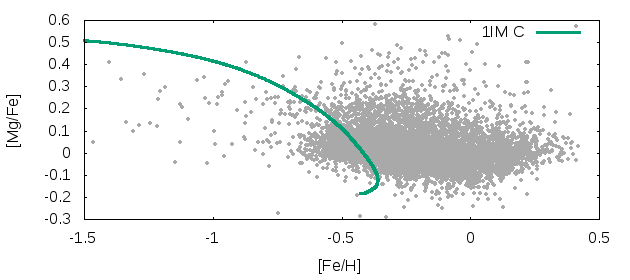}
\includegraphics[scale=0.35]{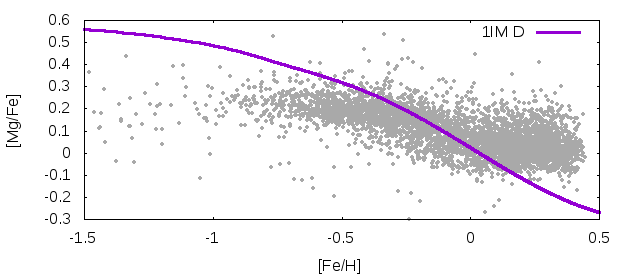}
\includegraphics[scale=0.35]{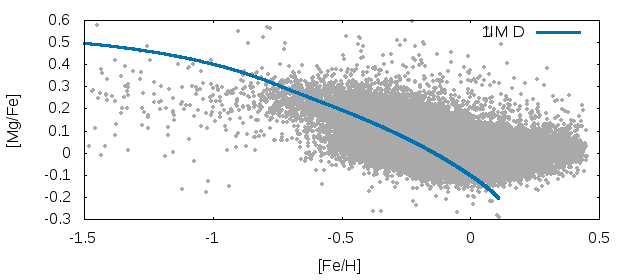}
\includegraphics[scale=0.35]{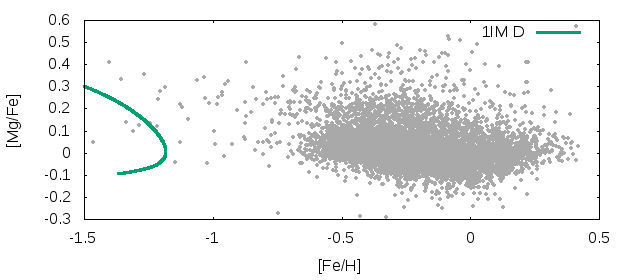}
 \caption{Observed and predicted abundance patterns of [Mg/Fe] vs. [Fe/H]. The data are from Hayden et al. (2015) and are divided as follows: 3-7 kpc (left panels), 7-11 kpc (central panels), 11-15 kpc (right panels). The predictions are from model 2IM A (upper panels), 1IM A (second panels), 1IM B (third panels), 1IM C (fourth panels), 1IM D (fifth panels) at different Galactocentric distances: 4 kpc (magenta line), 8 (blue line), 14 (green line).}
 \label{fig_02}
\end{figure*}

\section{Model results}

We consider the chemical evolution models for the Galactic disc developed by Grisoni et al. (2017) and we study the radial abundance gradients along the Galactic thin disc and its dependence upon several parameters: i) the timescale for the formation of the thin disc, increasing with Galactic radius according to the inside-out scenario (Matteucci and Francois 1989; Chiappini et al. 2001); ii) the SFE of the thin disc (Colavitti et al. 2009); iii) the radial gas flows, with a variable gas speed (Spitoni and Matteucci 2011).
\\In Table 1, we summarize the input parameters of the chemical evolution models. In the first column, we write the name of the model. In the second column, there is the SFE of the thin disc at different radii (4-6-8-10-12-14-16 kpc from the Galactic center). Finally, in the last column, we indicate the presence or the absence of radial gas flows. In each model, we adopt Kroupa et al. (1993) IMF and the inside-out law for the timescale of mass accretion in the thin disc, as expressed by Eq. \,\eqref{eq_02a}. 2IM A is the two-infall model with inside-out scenario for the Galactic thin disc. Similarly, 1IM A is the one-infall model for the thin disc with inside-out scenario. 1IM B is the one-infall model for the Galactic thin disc with inside-out and also a variable SFE. 1IM C is the one-infall model for the Galactic thin disc with inside-out and also the implementation of radial gas flows. 1IM D is the one-infall model for the Galactic disc with inside-out and both a variable SFE and radial gas flows.
\\Before discussing the abundance patterns and gradients for these models, in Fig. \ref{fig_01} we show the time evolution of the SFR as predicted by the 2IM A at various Galactocentric distance: 4, 6, 8, 10, 12, 14 kpc. The SFR during the thick disc phase is the same for every Galactocentric distance up to 8 kpc, because the assumed thick disc mass density in this model is constant up to 8 kpc. However, for R>8 kpc the thick disc mass density is assumed to go with the inverse of the distance and this is reflected into the SFR which is damped at large Galactocentric distances. We note that there is no real gap between the thick and thin disc phases due to the fact that there is no assumed threshold for the star formation, but still there is a quenching of star formation between the two phases. In the thin disc phase, the SFR is much higher at smaller Galactocentric distances since the total surface mass density is higher (see Eq.\,\eqref{eq_02a}).
\\In the following, we show the results for the abundance patterns ([Mg/Fe] vs. [Fe/H]) at various Galactocentric distances, for the present-day gradients and for the time evolution of gradients.

\begin{figure*}
\includegraphics[scale=0.6]{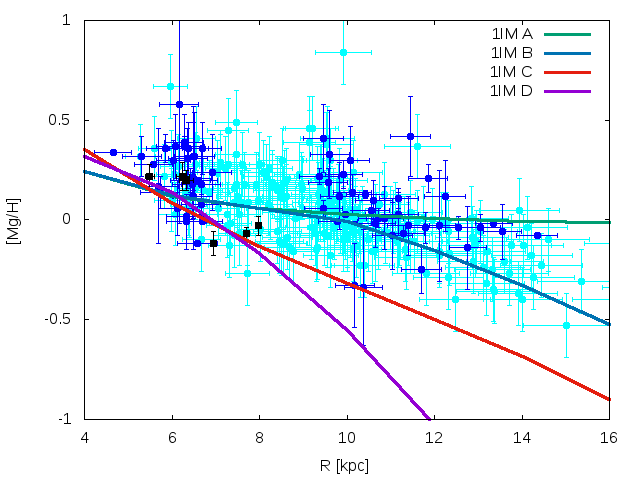}
 \caption{Observed and predicted radial abundance gradient for magnesium from Cepheids and young open clusters. The data are from Luck and Lambert 2011 (light-blue dots) and Genovali et al. 2015 (blue dots) for Cepheids, and from Magrini et al. 2017 (black squares) for young open clusters. The predictions are from model 1IM A (green line), 1IM B (blue line), 1IM C (red line) and 1IM D (magenta line).}
 \label{fig_03}
\end{figure*}

\subsection{Abundance patterns}

Our starting point is represented by the study of Grisoni et al. (2017), which explored the two different scenarios for the Galactic thick and thin discs in the solar neighborhood: the two-infall and the parallel one. In Fig. \ref{fig_01b}, we report the predictions of the two different theoretical approaches, i.e. the two-infall and the parallel models. In the case of the two-infall model (left panel), we predict an overabundance of Mg relative to Fe almost constant until [Fe/H]$<$-1.5 dex and then for [Fe/H]$\sim$-1.5 dex the trend shows a decrease due to the delayed explosion of Type Ia SNe. This behaviour of the abundance patterns of $\alpha$-elements such as Mg is well-interpreted in terms of the time-delay model due to the delay of iron ejection from Type Ia SNe relative to the faster production of $\alpha$-elements by core-collapse SNe (see Matteucci 2001; 2012). Then, there is a gap, which marks the transition between the thick and thin disc phases, and then the thin disc phase starts. On the other hand, in the case of the parallel model (right panel), we have two distict evolutionary paths for the thick and thin discs, with the thick disc being more $\alpha$-enhanced due to the faster timescale of formation (see Grisoni et al. 2017 for further discussion on the two approaches; Grand et al. 2018 for an interpretation of the two sequences in the [$\alpha$/Fe] vs. [Fe/H] plane in terms of cosmological simulations).
\\Now, we focus on how the tracks in the abundance patterns vary with the Galactocentric distance and we compare our model predictions with APOGEE data (Hayden et al. 2015).
In Fig. \ref{fig_02}, we show the abundance patterns of [Mg/Fe] vs. [Fe/H] at different radii in the case of the models of Table 1.
In the upper panels, we show the predictions of the two-infall model 2IM A at various Galactocentric distances. We note that the various tracks overlap in the thick disc phase, because we do not assume an inside-out formation for the thick disc (see also Haywood et al. 2018). Then, there is a dilution which is due to the second infall episode. In the thin disc phase, we can see that there is a slight difference between the various tracks due do the inside-out scenario for the thin disc, but the effect is not very noticeable. Similarly, for the 1IM A for the Galactic thin disc with only inside-out, the various tracks at different radii are almost overlapping and there is not so much difference between them. On the other hand, in the case of the 1IM B with variable SFE, the tracks are different, as the variable SFE increases the spread among them and the agreement with the data is good, in particular for the outer radii. Similarly, Spitoni et al. (2015) found that a variable SFE can reproduce the spread in the [$\alpha$/Fe] vs. [Fe/H] plot. Then, we show the results for the 1IM C with radial gas flows. Also in this case, we have a much larger spread among the various tracks due to the presence of radial gas flows, even if the effect is different with respect to the previous case. In fact, in the case of radial gas flows, the various tracks overlap at low metallicities, then radial gas flows become relevant for larger metallicities and the spread between the tracks appears. Finally, in the lowest panels of Fig. \ref{fig_02}, we show the results for the 1IM D with both variable SFE and radial gas flows. In this case, the spread among the various tracks is present both at low and high metallicities, due to the fact that we have combined the two effects: on one hand, the variable SFE, which acts at lower metallicities, and on the other, the radial gas flows, which act at higher metallicities. We can see that the combined effects are too strong, since the law for the variable SFE and the radial gas flows have been fine-tuned separately to best fit the data. Other combinations of the radial gas flows and variable efficiency of star formation, including the ones of Spitoni et al. (2015) (with constant radial gas flows and variable efficiency), have been tested and they produce results in between model 1IM C and 1IM D.
\\So, concluding, the best model to reproduce the APOGEE data is the model with a variable SFE, which can recover the spread among the various tracks and can provide a good match with observations.
\\We note that there is a clear discrepancy between all model predictions and observations at solar and super-solar metallicities. In fact, none of the models can reproduce [Mg/Fe] at high [Fe/H]. The observations indicate that [Mg/Fe] is essentially flat for [Fe/H]>0, while all models predict a decline. The disagreement between model and observations is probably related to the choice of the Mg stellar yields and/or to our poor understanding of the complete processes involved in the nucleosynthesis of Mg. For example, Romano et al. (2010) suggested the need for either a revision of current SNII and HN yields for solar and/or super-solar metallicity stars, or larger contributions to Mg productions from SNIa, or significant Mg synthesis in low and intermediate mass stars, or a combination of all these factors to reproduce the behavior of Mg. This has been extensively discussed in Magrini et al. (2017), where the differences between $\alpha$-elements are presented. In Grisoni et al. (2017), we also discussed the nature of the metal-rich high-alpha stars (MRHA) of Mikolaitis et al. (2017), which still have to be well-understood in terms of Galactic chemical evolution models. We concluded that in the parallel approach, the MRHA stars can be interpreted as metal-rich thick disc stars, whereas in the two-infall approach they can only be explained by invoking radial migration from the inner disc. Radial migration in Galactic discs is the subject of many investigations and it has been included in several chemical evolution models and simulations (Sch{\"o}nrich and Binney 2009; Minchev et al. 2013; Kubryk et al. 2015a,b; Spitoni et al. 2015; Grand et al. 2015).

\begin{figure*}
\includegraphics[scale=0.6]{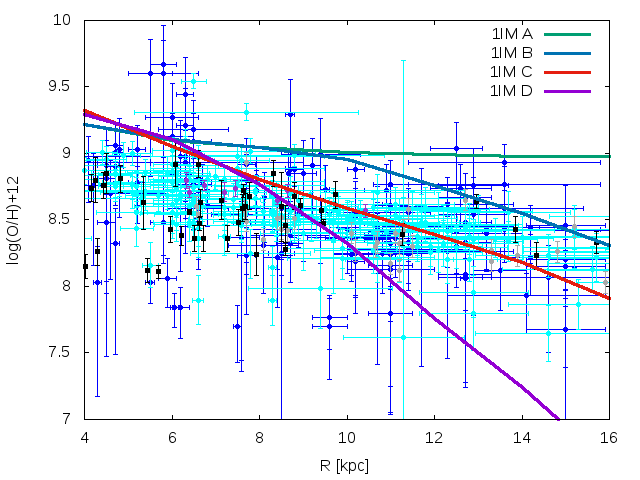}
 \caption{Observed and predicted radial abundance gradient for oxygen from HII regions and young planetary nebulae. The data are from Deharveng et al. 2000 (gray dots), Esteban et al. 2005 (violet dots), Rudolph et al. 2006 (blue dots), Balser et al. 2015 (light-blue dots) for HII regions, and from Stanghellini and Haywood 2018 (black squares) for young PNe. The predictions are from model 1IM A (green line), 1IM B (blue line), 1IM C (red line) and 1IM D (magenta line).}
 \label{fig_04}
\end{figure*}

\begin{figure*}
\includegraphics[scale=0.35]{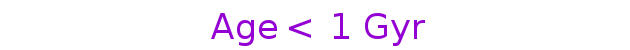}
\includegraphics[scale=0.35]{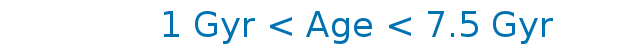}
\includegraphics[scale=0.35]{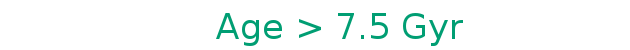}
\includegraphics[scale=0.35]{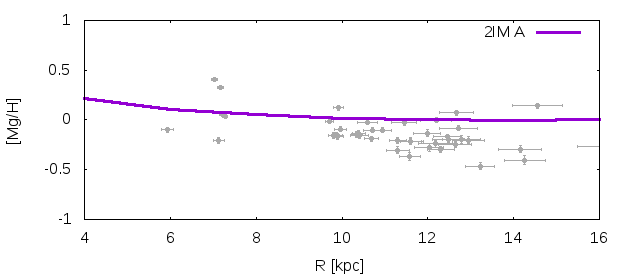}
\includegraphics[scale=0.35]{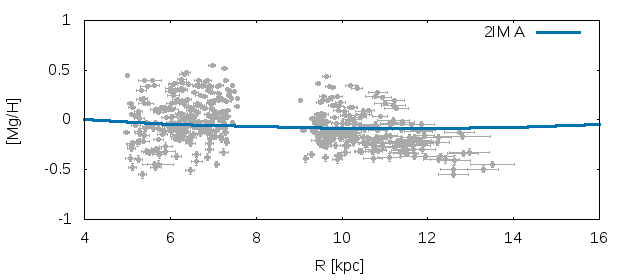}
\includegraphics[scale=0.35]{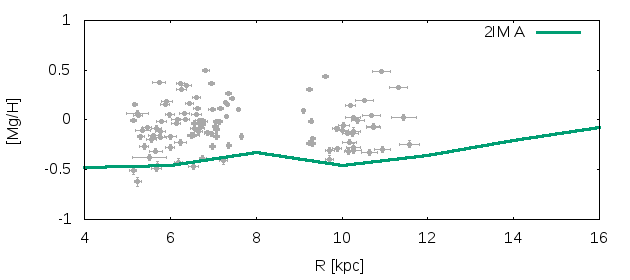}
\includegraphics[scale=0.35]{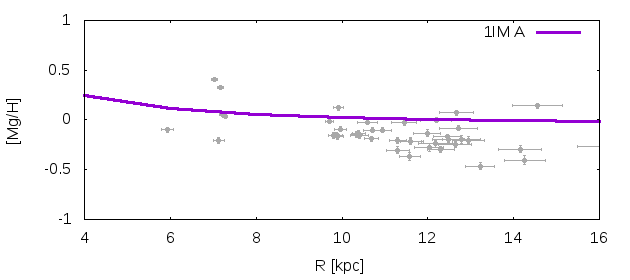}
\includegraphics[scale=0.35]{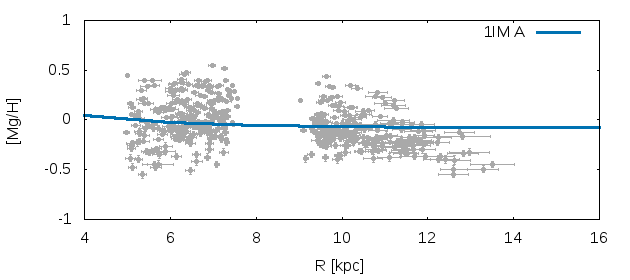}
\includegraphics[scale=0.35]{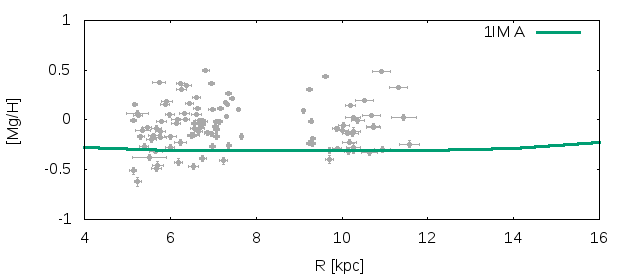}
\includegraphics[scale=0.35]{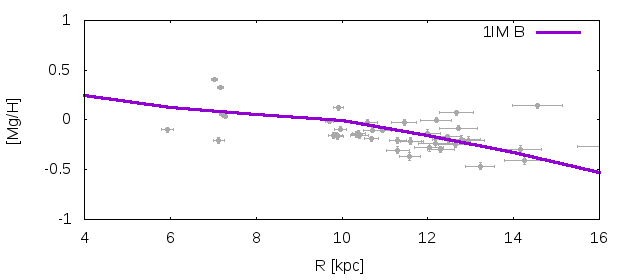}
\includegraphics[scale=0.35]{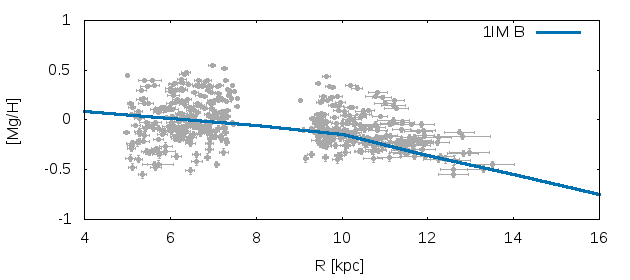}
\includegraphics[scale=0.35]{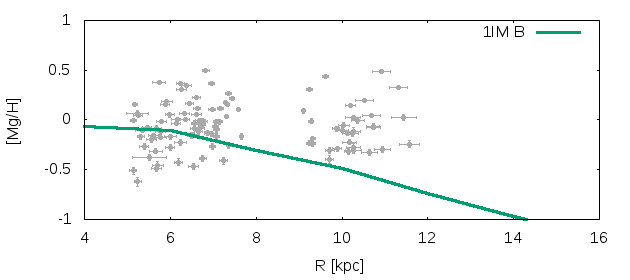}
\includegraphics[scale=0.35]{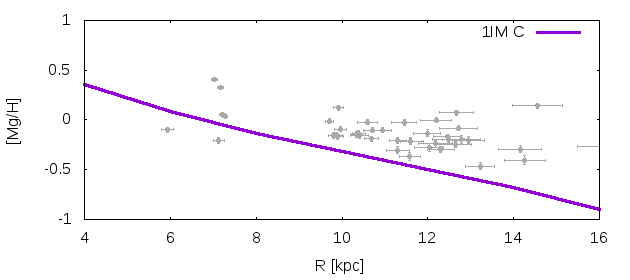}
\includegraphics[scale=0.35]{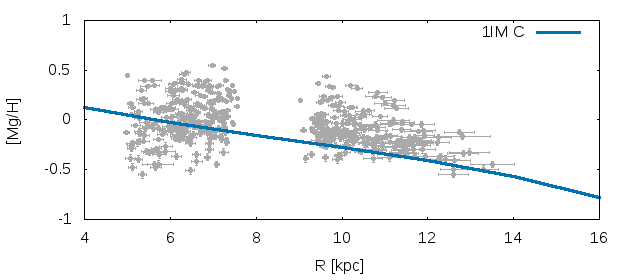}
\includegraphics[scale=0.35]{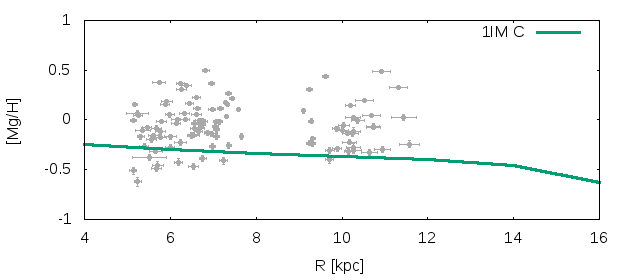}
\includegraphics[scale=0.35]{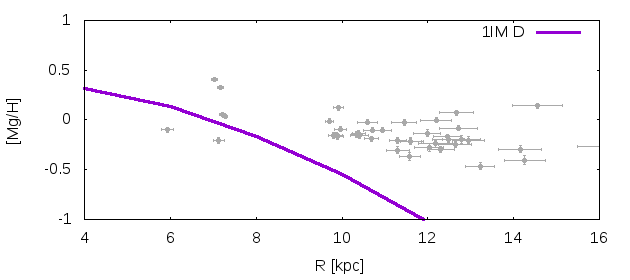}
\includegraphics[scale=0.35]{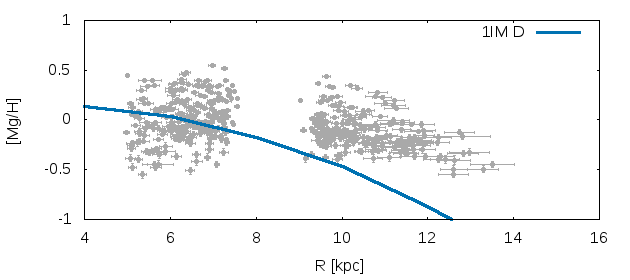}
\includegraphics[scale=0.35]{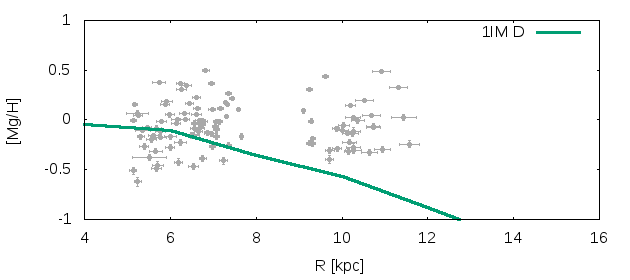}
 \caption{Time evolution of the radial abundance gradient for magnesium. The data are from Anders et al. (2017) and are divided in age bins as follows: younger than 1 Gyr (left panels), between 1 and 7.5 Gyr (central panels), older than 7.5 Gyr (right panels). The predictions are from model 2IM A (upper panels), 1IM A (second panels), 1IM B (third panels), 1IM C (fourth panels), 1IM D (fifth panels) at the various times.}
 \label{fig_05}
\end{figure*}

\begin{figure*}
\includegraphics[scale=0.36]{age1.png}
\includegraphics[scale=0.36]{age7.png}
\includegraphics[scale=0.36]{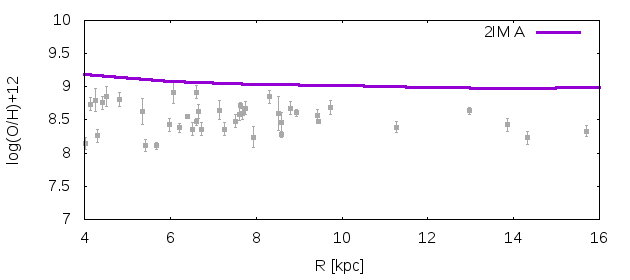}
\includegraphics[scale=0.36]{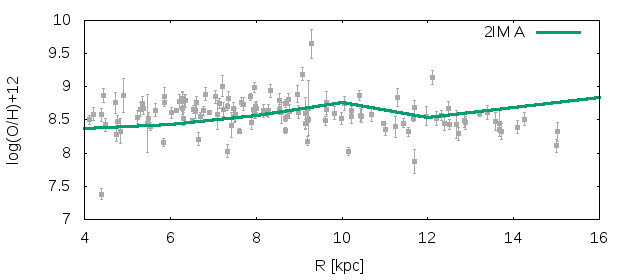}
\includegraphics[scale=0.36]{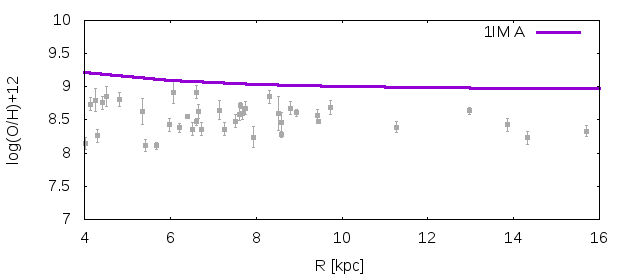}
\includegraphics[scale=0.36]{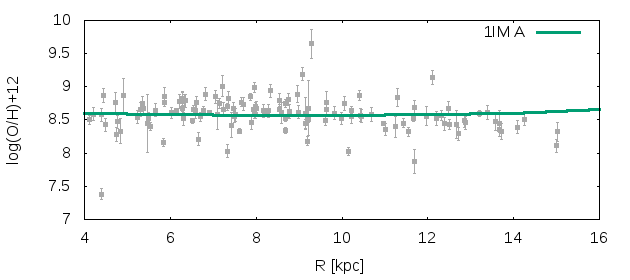}
\includegraphics[scale=0.36]{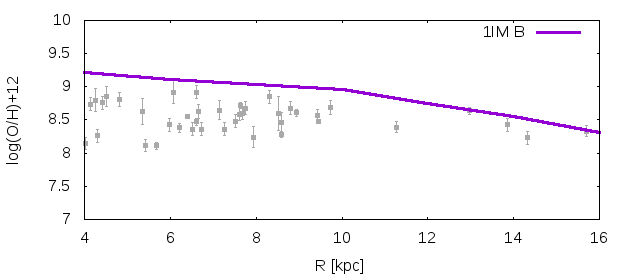}
\includegraphics[scale=0.36]{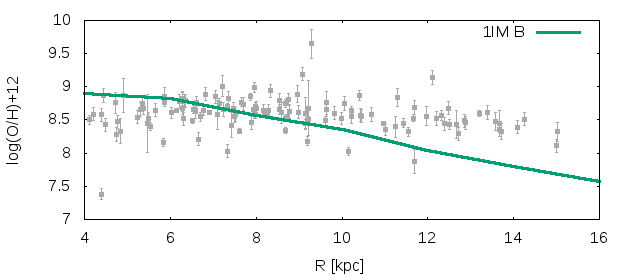}
\includegraphics[scale=0.36]{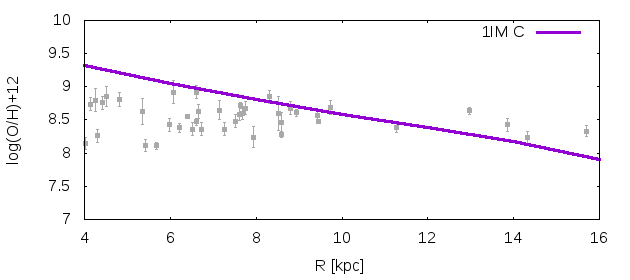}
\includegraphics[scale=0.36]{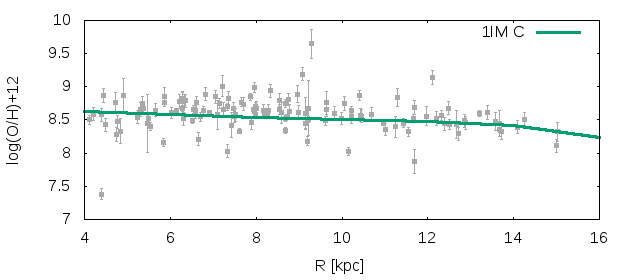}
\includegraphics[scale=0.36]{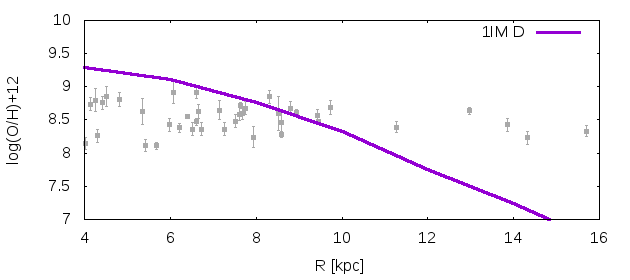}
\includegraphics[scale=0.36]{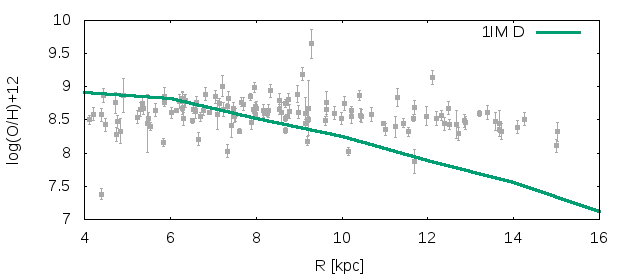}
 \caption{Time evolution of the radial abundance gradient for oxygen. The data are from Stanghellini and Haywood (2018) and are divided in age bins as follows: younger than 1 Gyr (left panels) and older than 7.5 Gyr (right panels). The predictions are from model 2IM A (upper panels), 1IM A (second panels), 1IM B (third panels), 1IM C (fourth panels), 1IM D (fifth panels) at the various times.}
 \label{fig_06}
\end{figure*}

\subsection{Present-day gradients}

Here, we consider the present-day abundance gradients.
\\In Fig. \ref{fig_03}, we show the observed and predicted radial abundance gradient for Mg from Cepheids and OCs. Cepheids are an essentially young populations, whereas OCs can have different ages and so not all OCs are tracers of the present time gradient; thus, here we consider only the young OCs of Magrini et al. (2017), with ages less than 1 Gyr. The predictions are from the one-infall model for the Galactic thin disc in the four different cases summarized in Table 1: 1IM A with only inside-out (light-blue line), 1IM B with variable SFE (blue line), 1IM C with radial flows (red line) and 1IM D (magenta line). We can see that the model with only inside-out scenario predicts a very flat gradient, whereas to steepen the gradients we need more ingredients such as a variable SFE or radial gas flows, and by combining the two ingredients we get an even steeper gradient. As we have noted in the case of abundance patterns at different Galactocentric distances, only inside-out does not seem to be sufficient to explain entirely the observations.
\\In Fig. \ref{fig_04}, we show the observed and predicted radial abundance gradient for oxygen from HII regions and PNe. Only young PNe can be considered together with HII regions, whereas older PNe are not tracers of the present time radial O/H gradient; thus, here we consider only YYPNe, i.e. PNe whose progenitor stars are younger than 1 Gyr (Stanghellini et al. 2018). The predictions are from the same models as in the previous figure, but compared with data from HII regions and young PNe. Also in this case and even more evidently, the model with only inside-out is not sufficient to explain the steep abundance gradient: we need a variable SFE or radial flows to explain the observational data. Therefore, a variable SFE or radial flows are important ingredients for obtaining a steeper gradient (see also Spitoni and Matteucci 2011).
\\These conclusions are in agreement with the comparison with APOGEE data for abundance ratios, for which we could not recover the spread with only inside-out, but we had to add the variable SFE and radial flows to better fit the observations.
In this context, radial migration can also have an effect, as already discussed at the end of the previous section. For example, Loebman et al. (2016) claimed that the results of radial migration can be important. If migration is important in well defined parts of the Galaxy, as for instance the outer disc, it can change the shape of the gradient. In particular, radial migration should flatten the gradient, but it has been shown that the effect may not be large for stars in the Milky Way (Di Matteo et al. 2013; Kubryk et al. 2013; Bovy et al. 2014).

\subsection{Time evolution of the gradients}

We focus here on the time evolution of the abundance gradients and show how abundance gradients vary with time in the various scenarios and we compare our model predictions with recent observational data on the time evolution of the radial metallicity gradients (Anders et al. 2017; Stanghellini and Haywood 2018).
\\In Fig. \ref{fig_05}, we show the time evolution of the radial abundance gradient for Mg. The predictions are from models 2IM A, 1IM A, 1IM B, 1IM C and 1IM D at different times (2 Gyr, 7 Gyr and 13.6 Gyr).
In the first panels, we show the predictions of the 2IM A. We can see that the two-infall model predicts a gradient inversion at early times: the gradient has a positive slope (at 2 Gyr), and then it flattens and reaches a slightly negative slope (at 13.6 Gyr). This gradient inversion was observationally claimed by Cresci et al. (2010), who have found an inversion of the O gradient at redshift z=3 in some Lyman-break galaxies. They showed that the O abundance decreases going toward the Galactic center, thus producing a positive gradient. This inversion was already noted theoretically by Chiappini et al. (2001) and studied by Mott et al. (2013) in terms of the two-infall model. This is a characteristic feature of the two-infall model, with its second infall episode of primordial gas which dilutes the gas in the inner regions in spite of the chemical enrichment.
In the second panels, we show the predictions of the 1IM A. At variance with the previous case, in the one-infall approach we have no gradient inversion due to the fact that we have no second infall episode of primordial gas which provokes the dilution. The slope of the gradient slightly steepens with time, but the effects is not noticeable, since in the 1IM A we assume only inside-out.
In the third panels, we show the predictions of the 1IM B with inside-out and also variable SFE. In this case, the time evolution of the gradient is different than in the previous case and the gradient flattens in time. This result is in agreement with other studies which predict a flattening of the gradient with time (Prantzos and Boissier 2000; Moll{\'a} and \& D{\'{\i}}az 2005; Vincenzo and Kobayashi 2018; Minchev et al. 2018).
In the fourth panels, we show the predictions of 1IM C with inside-out and also radial gas flows. The models with radial gas flows predict that the gradient steepens noticeably in time as was found by Mott et al. (2013), but then we have no gradient inversion because it is a one-infall model and not a two-infall one.
Finally, in the last panels, we show the predictions of the 1IM D with inside-out and also both the variable SFE and radial gas flows. We can see that the gradient starts steep due to the variable SFE at different distances and then it becomes even steeper in time due to the effect of radial gas flows, which becomes dominant at later time.
\\Then, in Fig. \ref{fig_06}, we compare our model predictions also with PNe data from Stanghellini and Haywood (2018), which show that the OPPNe oxygen gradient is shallower than that derived from YPPNe. Also in this case, the predictions are from models 2IM A, 1IM A, 1IM B, 1IM C and 1IM D at the different times.
\\Concluding, the model 1IMB with the variable SFE still provides a good agreement with the observational data, in particular at recent times, but it predicts a steeper behaviour at earlier times which is not present in the data. To recover a flatter gradient at earlier times, we should rather consider models 1IMA and 1IMC, whereas with the two-infall model 2IMA we can even get an evident gradient inversion, as we have discussed previously. The difference between model predictions is due to the fact that the gas chemical evolution is very sensitive to the prescriptions of the physical processes that lead to the enrichment of inner and outer discs, and the flattening or steepening of gradients in time depends on the interplay between infall rate and star formation rate along the Galactic disc. Different recipes of the star formation process or gas accretion mechanisms can provide very different predictions for the abundance gradients.

\section{Conclusions}

In this paper, we have studied the formation and chemical evolution of the Milky Way discs with particular focus on the abundance patterns at different Galactocentric distances, the present-time abundance gradients along the disc and the time evolution of abundance gradients. We have considered the recently developed chemical evolution models by Grisoni et al. (2017) for the solar neighborhood, both the two-infall and the one-infall, and we have extended our analysis to the other Galactocentric distances, also implementing radial gas flows in the code for this purpose. In particular, we have examined the processes which mainly influence the formation of abundance gradients: i) the inside-out scenario for the formation of the Galactic thin disc, ii) a variable star formation efficiency, and iii) radial gas flows along the Galactic disc.
\\Our main conclusions are as follows.
\begin{itemize}
\item As regard to the abundance patterns (in particular [Mg/Fe] vs. [Fe/H]) at different Galactocentric distances, the inside-out scenario for the thin disc is a key element, but provides only a slight difference between the various tracks at different radii and so it is not sufficient to explain the data at various radii. In order to have a more significant spread among the various tracks, we need further ingredients such as a variable star formation efficiency or radial gas flows: the variable star formation efficiency produces a spread at lower metallicities, wheres the radial gas flows become significant at higher ones. The case with a variable star formation efficiency provides a very good agreement with the observational data, in particular for the outer radii. However, we note that none of the models can reproduce [Mg/Fe] at high [Fe/H], and this can be due to a general problem in our understanding of Mg production (as also pointed out by Romano et al. 2010; Magrini et al. 2017).
\item Also concerning the present-day abundance gradients along the Galactic thin disc, the inside-out scenario provides a too flat gradient and cannot explain the observational data, neither of Cepheids, young OCs, young PNe and HII regions which show a steeper gradient. To recover the steeper gradient, we need the variable star formation efficiency or radial gas flows. This conclusion is in agreement with the comparison with the data for abundance ratios.
\item On the other hand, for the time evolution of abundance gradients, the model with the variable star formation efficiency provides a good agreement with the observational data at recent times, but it predicts a steeper behaviour at earlier times which is not present in the data. To reproduce a flatter gradient at earlier times, we should rather consider the models with constant star formation efficiency or we would need radial migration, more efficient for the older populations. Thus, what we are observing is a gradient flattened by radial migration, and not the original one (see for instance Magrini et al. 2016). With the two-infall model, we can even get an evident gradient inversion at high redshift, when the efficiency of star formation is constant.
\item In our scenario, the Galactic thick disc formed on a very short timescale ($\tau_1=0.1$ Gyr, see Grisoni et al. 2017), which is assumed to be constant with radius. Therefore, there is no inside-out scenario for the thick disc, in agreement with Haywood et al. (2018).
\end{itemize}
In summary, we conclude that the inside-out scenario is a key ingredient for the formation of Galactic discs, but cannot be the only one to explain abundance patterns at different Galactocentric distances and abundance gradients. Further ingredients are needed, such as a variable star formation efficiency and radial gas flows; also radial migration could have an effect, although it has been shown that this may not be a large factor for stars in the Milky Way (Di Matteo et al. 2013; Kubryk et al. 2013; Bovy et al. 2014). Our model with a variable star formation efficiency is in very good agreement with the observational data, both the abundance patterns at different Galactocentric distances and abundance gradients, even if it predicts a too steep behaviour at earlier times. The flattening or steepening of gradients in time is due to the fact that the gas chemical evolution is very sensitive to the prescriptions of the physical processes that lead to the enrichment of inner and outer discs, mainly to the constancy or variability of the star formation efficiency. Therefore, different recipes of the star formation process or gas accretion mechanisms can provide very different predictions for the abundance gradients, as we have shown in this work.

\section*{Acknowledgments}

V.G., E.S. and F.M. acknowledge financial support from the University of Trieste (FRA2016). The authors thank the referee for the useful comments and suggestions, which have certainly improved the paper.

\end{document}